\documentclass[a4paper,amsmath,amssymb]{jpconf}
\usepackage{graphicx}
\usepackage{amsmath, amsthm, amssymb}
\begin{document}
\title{Polar vs. apolar alignment in systems of polar self-propelled particles}

\author{Fernando Peruani}
\address{Max Planck for the Physics of Complex Systems, N\"othnitzer str. 38,
  Dresden, Germany}
\ead{peruani@pks.mpg.de}

\author{Francesco Ginelli}
\address{Service de Physique de l'Etat Condens\'e, CEA-Saclay, 91191 Gif-sur-Yvette, France}

\author{Markus B\"ar}
\address{Physikalisch-Technische Bundesanstalt, Abbestrasse 2-12, 10587
  Berlin, Germany}

\author{Hugues Chat\'e}
\address{Service de Physique de l'Etat Condens\'e, CEA-Saclay, 91191 Gif-sur-Yvette, France}

\begin{abstract}
The symmetry of the alignment mechanism in systems of polar self-propelled particles 
determines the possible macroscopic
large-scale patterns that can emerge.
Here we compare polar and apolar alignment. 
These systems share some common features like giant number fluctuations in the
ordered phase and self-segregation in the form of bands near the onset of
orientational order. 
Despite these similarities, there are essential differences like the symmetry
of the ordered phase and the stability of the bands. 
\end{abstract}
\section{Introduction}
\label{sec:Introduction}

Systems of collectively moving entities are ubiquitous in nature,  ranging from  
 flocks of birds~\cite{animals,cavagna10,bhattacharya10}, 
insect swarms~\cite{buhl06,romanczuk09}, bacterial collective motion~\cite{zhang10,peruani2011}, to even driven granular media~\cite{narayan07,kudrolli08,kudrolli10,deseigne10}. 
Beyond the complexity of each system, it is possible to classify the interaction among the moving entities according to its symmetry, which can be either polar or apolar. 
Polar (or ferromagnetic) interactions lead to  parallel alignment of the velocity of the moving objects, while apolar (or nematic) interactions allow both, parallel as well as antiparallel alignment. 
Fish, birds, and insects seem to exhibit polar alignment~\cite{animals,cavagna10,bhattacharya10,buhl06,romanczuk09}. 
Bacteria and driven granular media often display apolar alignment~\cite{peruani2011,narayan07,kudrolli08,kudrolli10,deseigne10}.  
The physical origin of the alignment mechanism varies among these examples and it is difficult to identify a particular physical interaction with a given symmetry in these non-equilibrium systems. 
For instance, steric interactions among elongated self-propelled objects lead to apolar alignment. This can be easily observed in realistic self-propelled rod  models~\cite{peruani06}, experiments with driven rods~\cite{kudrolli08,kudrolli10}, and experiments with gliding bacteria~\cite{peruani2011}. 
However, if the objects are isotropic, volume exclusion effects can result in a polar alignment, as observed in models~\cite{grossman2008} and experiments~\cite{deseigne10} with polarly driven disks. 

Here we focus on the collective large-scale patterns emerging in systems of point particles moving in two-dimensions at constant speed. 
We compare the dynamics and macroscopic properties of self-propelled particles interacting by a polar and an apolar velocity alignment mechanism. 
We list the common features and highlight the differences. 
We provide in this way a summary of the two main classes of polar self-propelled particle systems.

\section{Equation of motion of self-propelled particles}
The evolution of the $i$-th particle is given by
the following updating rules:
\begin{eqnarray}\label{motion_pos}
\mathbf{x}_{i}^{t+\Delta t }&=&\mathbf{x}_{i}^{t} +v_0 e^{\text{i}\,\theta_i^{t}} \Delta t \\
\label{motion_vel} \theta_i^{t+\Delta t }
&=&\arg\left(\sum_{\left|\mathbf{x}_{i}^{t}-\mathbf{x}_{j}^{t}\right|\leq\epsilon}
  f \left( \theta_j^{t},\theta_i^{t}\right) e^{\text{i}\,\theta_j^{t}}  \right)+\eta_{i}^{t}
\end{eqnarray}
where $\mathbf{x}_{i}^{t}$ is the position of the particle and $\theta_i^{t}$ its  direction of motion at time $t$, $v_0$ the active particle
speed, $\arg\left(\mathbf{b}\right)$
indicates the argument of the imaginary number $\mathbf{b}$ ,  $\eta_{i}^{t}$ is a delta-correlated white noise of strength $\eta$, and
$\Delta t$ is the temporal time step. 
Notice that Eqs. (\ref{motion_pos}) and (\ref{motion_vel}) can be considered the limiting
case of very fast angular relaxation of a system of equations of the form: 
\begin{eqnarray}\nonumber 
\begin{array}{cc}
\dot{\mathbf{x}}_i= v_0 e^{\text{i}\,\theta_i}\,, &  \dot{\theta_i}= - \gamma \frac{\partial
U}{\partial \theta_i}(\mathbf{x_i},\theta_i) +\tilde{\eta}_{i}(t)\,,\\
\end{array}
\end{eqnarray}
 as discussed in~\cite{peruani08}. 
The symmetry of the alignment mechanism is contained in the function $f$ which
is defined as:
\begin{equation}\label{eq:f}
f\left( \theta_j^{t},\theta_i^{t}\right) = 
\begin{cases} 
1 & \mbox{for polar  alignment} \\
\mbox{sign}\left(\cos( \theta_j^{t}-\theta_i^{t})\right) & \mbox{for apolar alignment} \\
\end{cases} \, ,
\end{equation}
where $\mbox{sign}\left(x\right)$ return a $+1$ if $x\geq0$ and $-1$ otherwise. 
Eqs. (\ref{motion_pos}) and (\ref{motion_vel}) together with Eq.~(\ref{eq:f})
defined the so-called Vicsek model~\cite{vicsek95,chate08} for polar alignment,
and the model for (ideal) self-propelled rods introduced
in~\cite{peruani08,ginelli10} for apolar alignment.

\subsection{Order parameters}

The orientational order can be characterized by the following order parameters. 
The polar (ferromagnetic) order parameter is defined by:
\begin{eqnarray}\label{eq:orderparam_f}
\phi = \langle \left| \frac{1}{N} \sum_{k=1}^{N} \exp (\text{i}\, \theta^t_k)   
\right| \rangle \, ,
\end{eqnarray}
where $\langle \hdots \rangle$ denotes time-average and $N$ stands for the total number of particles in the system.
$\phi$ takes the
value $1$ when all particles move in the same direction, while in the
disordered phase, i.e., when particles move in any direction with equal
probability, it vanishes. 
On the other hand, the apolar (nematic) ordered parameter takes the form: 
\begin{eqnarray}\label{eq:orderparam_lc}
S = \langle \left| \frac{1}{N} \sum_{k=1}^{N} \exp (\text{i}\,2\, \theta^t_k)  
\right| \rangle \, . 
\end{eqnarray}
Formally, $S$ can be derived from the order parameter matrix $Q$ of {\it liquid
crystals} (LC)~\cite{doi}, as the largest eigenvalue (which here we have normalized
such that $S^{LC} \in [0,1]$).
When the system is perfectly nematically ordered, i.e., when particles move in
opposite directions along the same axis, $S$ takes the value $1$. 
In summary, a perfectly polarly ordered phase is 
characterized by $\phi=S=1$, while for a genuine apolarly  ordered
phase, $\phi=0$ and $S=1$.

\begin{figure*}[!t] 
\begin{center} 
\includegraphics[clip,width=12cm]{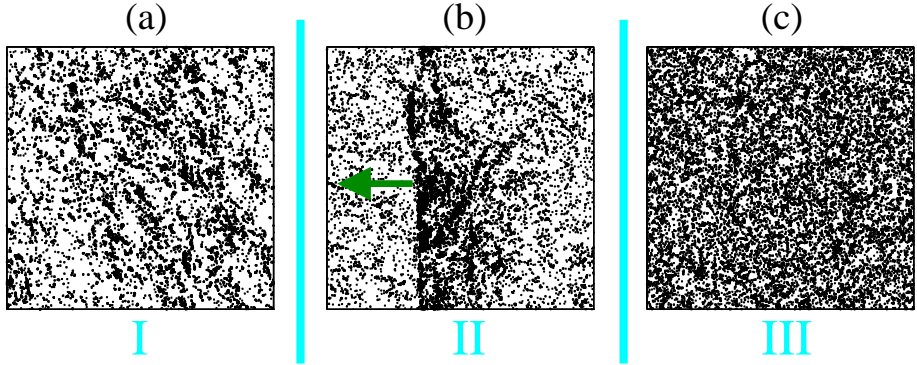} 
\caption{ 
Typical steady-state snapshots for SPP with polar alignment at different noise values 
(linear size $L=512$, particle density $\rho=1/8$, active speed  $v_{\rm 0}=1/4$). The snapshots correspond to phase I,
(a) $\eta=0.06$, phase II, (b) $\eta=0.1$, and phase III,  (c)
$\eta=0.19$. The arrow in (b) indicates the direction of motion of the
high-density band.}.
\label{fig:PolarSnapShots} 
\end{center} 
\end{figure*}

\section{Macroscopic patterns with polar alignment}

We start out by reviewing the large-scale properties of SPP with polar
alignment. 
If we fix the active speed $v_{\rm 0}$ and the density $\rho$, while varying
the 
noise strength $\eta$ we observe the emergence of three statistically distinct stationary
states or phases. 
Two of these phases correspond to orientationally ordered phases, while the third one is a
disordered phase. 
We refer to these phases as phase I, II, and III by increasing value of
$\eta$. 

Phase I exhibits polar long-range order and 
 is characterized by the absence of regular large-scale, high density
structures, see Fig.\ref{fig:PolarSnapShots}(a). 
Due to this fact, it is often said that this phase is spatially homogeneous
albeit with large density fluctuations. 
The presence of polar long-range order implies that $\phi(\eta) \to K(\eta)$
as the system size $L \to \infty$, with $K(\eta)>0$ a constant that depends
only on $\eta$.  
Phase I also exhibits giant number fluctuations (NF). NF are defined as $\Delta n^2(\ell)=\langle (n(\ell)-\langle
n\rangle(\ell))^2\rangle$, where $n(\ell)$ stands for the number of particles
in a box of linear size $\ell$. 
The average number of particles is $\langle n \rangle = \rho \ell^2$ 
while $\Delta n(\ell)$ is expected to be $\Delta n \propto \langle n \rangle
^\alpha$. 
Giant NF correspond to $\alpha>\frac{1}{2}$~\cite{ramaswamy03}. In~\cite{chate08} it was shown that $\alpha \sim 0.8$ for phase I.

By increasing $\eta$ we reach a point at which large-scale, elongated,
high-density, high-order solitary structures emerge.
These structures, that we refer to as bands, move at roughly constant speed, see Fig.\ref{fig:PolarSnapShots}(b). 
A system of SPPs with polar alignment can display multiple bands. 
The low-density region in between bands is disordered  
and 
there is no characteristic separation length between bands~\cite{chate08}. 
In summary, phase II is characterized by polar long-range order and the
presence of travelling bands. 
If $\eta$ is increased further, we reach phase III that exhibits local and
global disorder, see Fig.\ref{fig:PolarSnapShots}(c).

\begin{figure} 
\begin{center} 
\includegraphics[clip,width=8.6cm]{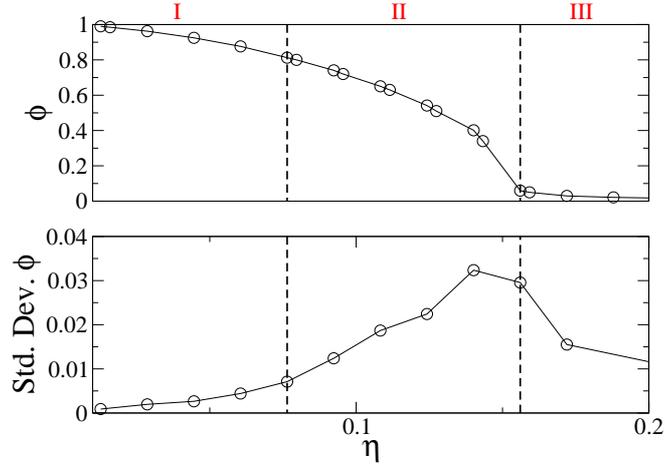}
\caption{Polar order parameter $\phi$ and its standard deviation as function
  of the noise amplitude $\eta$ (other parameters as in Fig. \ref{fig:PolarSnapShots}).} 
\label{fig:OrientationalOrder} 
\end{center} 
\end{figure} 

\section{Macroscopic patterns with apolar alignment}

\begin{figure*}[t] 
\begin{center} 
\includegraphics[clip,width=18cm]{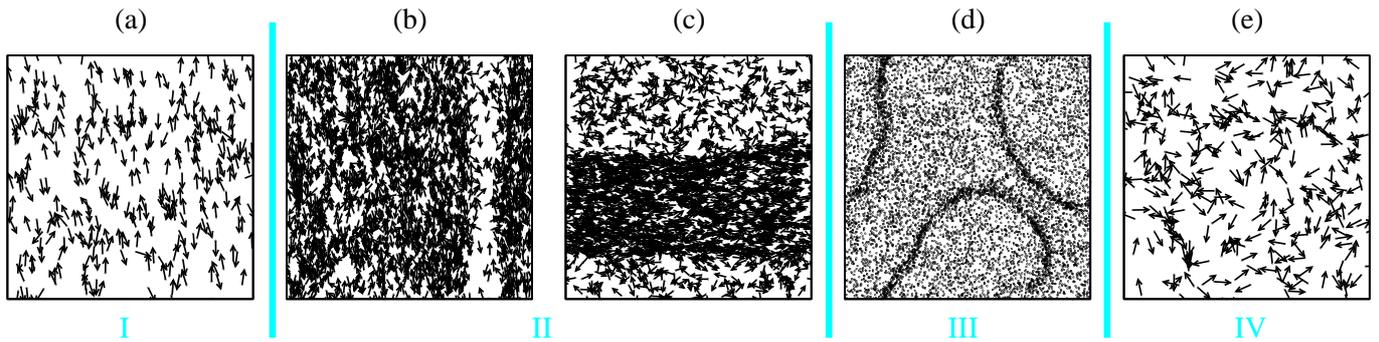} 
\caption{ 
Typical steady-state snapshots for SPP with apolar alignment at different noise values 
(linear size $L=2048$, density $\rho=1/8$, and velocity $v_0=1/2$). 
(a) $\eta=0.08$, (b) $\eta=0.10$, (c) $\eta=0.13$, (d) $\eta=0.168$, (e) $\eta=0.20$. 
Arrows indicate the polar orientation of particles (except in (d)); only a fraction of  
the particles are shown for clarity reasons. }
\label{fig:2} 
\end{center} 
\end{figure*}

\begin{figure} 
\begin{center} 
\includegraphics[clip,width=8.6cm]{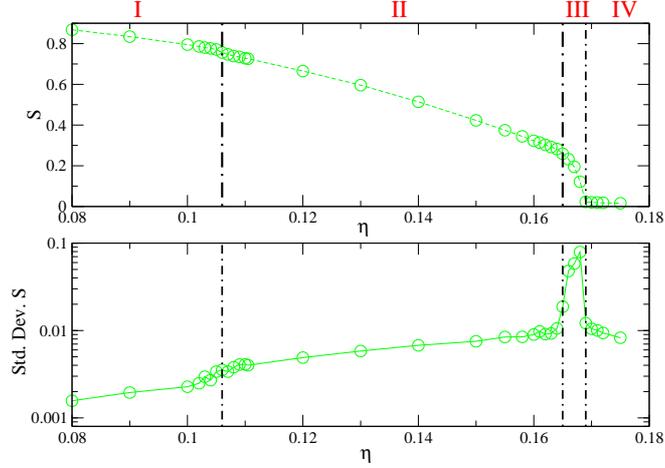}
\caption{Apolar (nematic) order parameter $S$  
and its standard deviation as function of the noise amplitude $\eta$ (other parameters as in Fig. \ref{fig:2}).}
\label{fig:3} 
\end{center} 
\end{figure}

\begin{figure}
\begin{center} 
\includegraphics[clip,width=8.6cm]{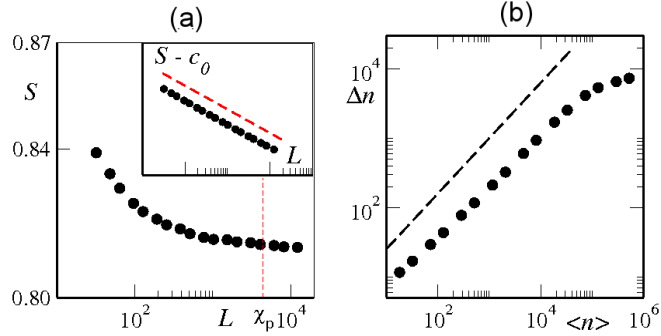} 
\caption{Phase I of SPPs with apolar alignment: long-range apolar order and giant number fluctuations ($\rho=1/8$ and $\eta=0.095$). 
(a) Apolar order parameter $S$ vs 
system size $L$ in square domains. The vertical red dashed line marks the persistence length  
$\chi \approx 4400$ that particles can travel before performing a U-turn.   
Inset: $S-C_0=0.813063$ vs $L$ (red dashed line: $L^{-2/3}$ decay). 
(b) Number fluctuations $\Delta n$ as function of $\langle n \rangle$. The
dashed line correspond to an algebraic growth with exponent $0.8$ (simulations
with $L=4096$). See Ref.~\cite{ginelli10}.} 
\label{fig:4} 
\end{center} 
\end{figure}

Now we focus on a system of SPP with apolar alignment.  
While sweeping $\eta$, keeping the active speed and the density fixed, we observe the emergence of four statistically distinct
stationary states or phases, see Figs.~\ref{fig:2}-\ref{fig:3}. 
Two of these four regimes correspond to ordered phases while the other two to
disordered phases. 
At the two extremes, very high and very low values of $\eta$, the system is
spatially homogeneous, Figs.~\ref{fig:2}(a,e).
At intermediate values of $\eta$, spontaneous density segregation occurs in
the form of a  high-density ordered region along which the particles move back
and forth, Figs.~\ref{fig:2}(b-d). 
It is important to stress that macroscopic polar order always remains near
zero. Nevertheless, if we look at very short length scales, we observe that
the apolar ordered phase is formed by polarly oriented clusters. 
In the following we provide a more quantitative description of these four phases, 
which we name, by increasing value of $\eta$, phase I to IV.
Phase I exhibits true long-range apolar (nematic) order and is spatially
homogeneous. 
This apolar order corresponds to two subpopulations of particles of roughly
equal size  
that migrate in opposite directions (Fig.~\ref{fig:4}a).
We conclude that the apolar order is truly long-range by looking at scaling of
the apolar order parameter $S$ with the system size. 
From Fig.~\ref{fig:4}a, it seems that $S(L)$ approaches a constant
asymptotic value $C_0(\eta)$ for $L\to\infty$, with $C_0(\eta)>0$, indicating the possible
existence of true long-range apolar order. 
Another remarkable feature of phase I is the presence of giant number
fluctuations (NF). Fig.~\ref{fig:4}b shows the scaling $\Delta n \propto
\langle n \rangle^{\alpha}$ in phase I, where it is found again that
$\alpha \sim 0.8$.

\begin{figure} 
\begin{center} 
\includegraphics[clip,width=8.6cm]{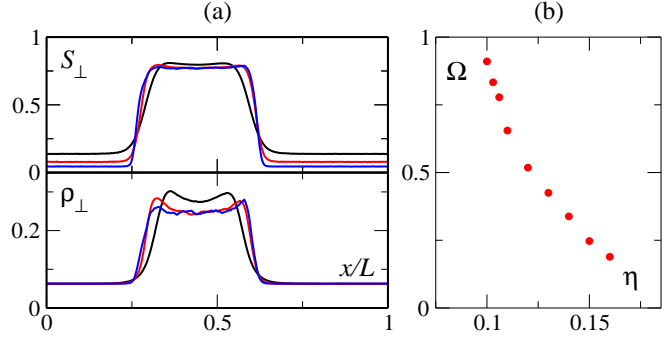} 
\caption{Phase II of SPPs with apolar alignment is characterized by the presence of a dense region. 
(a) Rescaled transverse profiles in square domains of linear size 
$L=512$ (black), $1024$ (red), and $2048$ (blue) at $\eta=0.14$.
(Data averaged over the longitudinal direction and time, translated to be centered
at the same location.)
Bottom: density profiles. Top: nematic order parameter profiles. 
(b) Surface fraction $\Omega$ as a function of noise amplitude $\eta$ (defined here as
the width at mid-height of the rescaled $S$ profile). See Ref.~\cite{ginelli10}.}
\label{fig:5} 
\end{center} 
\end{figure}

By increasing $\eta$ we move from phase I to phase II. 
Its onset is characterized by the emergence of a narrow, low-density, disordered channel.
This channel becomes wider at larger $\eta$ values, so that one can speak of a 
high-density ordered region along which particles travel in both directions. 
The high-density region exhibits apolar order with properties similar to those observed in
phase I: true long-range order and giant NF. 
The (rescaled) region possesses 
a well-defined profile with sharper and sharper edges as $L$ increases 
(Fig.~\ref{fig:5}a). The fraction area $\Omega$ occupied by the dense region is 
thus asymptotically independent of system size, and it decreases 
continuously as the noise strength $\eta$ increases (Fig.~\ref{fig:5}b). 
 
By increasing  $\eta$ even further, we reach phase III, where the dense region becomes unstable 
and constantly bend, merge, break and reform. 
As result of this dynamics, $S(t)$ fluctuates strongly and on very large time
scales, but its average decreases as $1/\sqrt{N}$~\cite{ginelli10}. Thus, phase III is
characterized by the absence of long-range order and the presence of large
correlation lengths and times. 
At even larger $\eta$ values, we find phase IV, which exhibits local and
global disorder on smal length- and time-scales, and is spatially homogeneous.

\begin{figure} 
\begin{center} 
\includegraphics[clip,width=8.6cm]{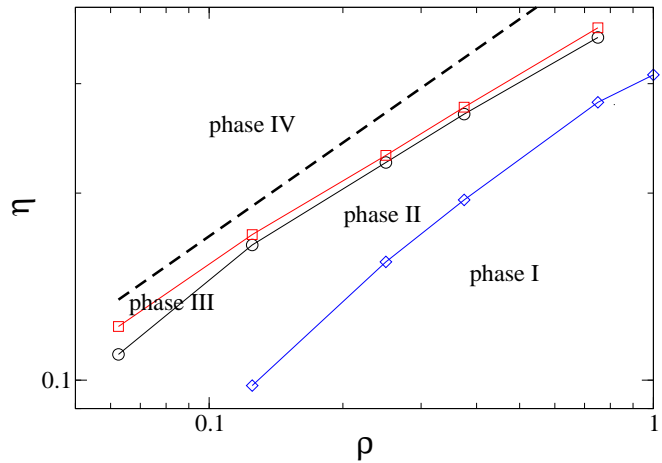} 
\caption{Phase diagram - Critical value of $\eta$, separating the four phases,
  as function of the density. Thresholds obtained for systems of size
  $L=1024$. The dashed line corresponds to the scaling $\eta \propto
  \rho^{1/2}$.  } 
\label{fig:phaseDiagram} 
\end{center} 
\end{figure}

\section{Comparison between polar and apolar alignment}

Self-propelled particles moving at constant speed with either polar or apolar
alignment exhibit a phase transition from a disordered to
orientationally ordered phase. 
For both alignments the order seems to be long-range, being purely polar for polar
alignment and purely apolar (nematic) for apolar alignment. 
It has been shown theoretically that self-propelled particles with polar
alignment can display long-range order~\cite{tonertu}. However, 
the claim about the existence of apolar long-range order is exclusively based on large-scale simulations
and a theoretical justification is still missing. 
Assuming that the findings here reported for large-scale simulations hold in
the thermodynamical limit, we conclude that SPPs with polar alignment exhibit
three phases, while with apolar alignment, the phases are four. 
For both alignments, there are two ordered phases, one of them spatially
``homogeneous'' though with giant number fluctuations with roughly the
same critical exponent $0.8$, while the other one 
is characterized by the presence of high-density, high ordered regions.
For polar alignment, the system can display multiple solitary bands. These bands are
travelling bands whose moving direction is perpendicular to the long axis of
the band, i.e., particles are aligned roughly in the same direction exhibited
by the band velocity.  
On the other hand, for apolar alignment, we observe only one dense region. This region
does not move and is composed of two populations of particles moving in
opposite direction along its periodic axis. 
For polar alignment there is one disordered phase characterized by local and
global disorder. SPPs with apolar alignment also exhibit a similar phase,
however these particles also display an intringuing disordered phase 
where particle self-segregation in the form of highly dynamical bands occur. 

This means that in a coarse-grained description of these systems, the
homogeneous disordered phase should get unstable by the emergence of  
soliton-like structures, the travelling bands, for polar alignment while for
apolar alignment, local alignment has to lead to the formation of highly
dynamical region. 
In the case of apolar alignment, the onset of global order is not associated
to the instability of the disorder homogeneous state, as for polar alignment,
but with the stability of the dense region. 
On the other extreme, well in the ``homogeneous'' ordered phase, we can approximate the
behavior of the order parameter $\phi$ and $S$ as:
\begin{eqnarray}\label{eq:sf_perfectorder}
\phi &=& \frac{2}{\eta} \sin(\eta/2)  \\
S &=&  \frac{1}{\eta^2}\sin^{2}(\eta) \, ,
\end{eqnarray}
using a simple mean-field argument as discussed in~\cite{peruani2010}. From this we can presume that for a
given set of parameters $v_0$, $\rho$, and $\eta$, $\phi(v_0, \rho, \eta) \geq
S(v_0, \rho, \eta)$. 
Moreover, a simple mean-field analysis of the``homogeneous'' disordered
phase~\cite{peruani08} 
reveals that, for fixed $v_0$ and $\rho$, this phase looses its stability at
lower values of $\eta$ for apolar alignment than for polar alignment.  
From all this, we learn that is easier to achieve orientational order with
polar than with apolar alignment. This comparison also suggests that polar
order is more robust than apolar order. 
Beyond this rough comparison between both alignment, a theoretical
understanding of the formation of dense regions in both systems is still lacking. 
For polar alignment, there are some  promising recent theoretical
results~\cite{mishra10,bertin} that provide an interesting perspective to
this puzzling issue. 
We hope that in the near future all these open problems will be fully
understood.



\end{document}